\documentclass[12pt,preprint]{aastex}

\shorttitle{A Massive Jet Ejection from SS~433}
\shortauthors{Kotani et al.}

\begin{document}

\title{A Massive Jet Ejection Event from the Microquasar SS~433
Accompanying Rapid X-Ray Variability}

\author{T. Kotani\altaffilmark{1}, S. A. Trushkin\altaffilmark{2},
R. Valiullin\altaffilmark{3},
K. Kinugasa\altaffilmark{4},
S. Safi-Harb\altaffilmark{5},
N. Kawai\altaffilmark{1}, and 
M. Namiki\altaffilmark{6}}
\email{kotani@hp.phys.titech.ac.jp}

\altaffiltext{1}{Tokyo Tech, 2-12-1 O-okayama, Tokyo 152-8551, Japan}
\altaffiltext{2} {Special Astrophysical Observatory RAS, Nizhnij Arkhyz,
Karachaevo-Cherkassia 369167, Russia}
\altaffiltext{3} {Astrophysical Institute of  Kazakh Academy  of
Sciences, 480020 Alma Ata, Kazakhstan}
\altaffiltext{4} {Gunma Astronomical Observatory, 6860-86 Nakayama,
Takayama, Agatsuma, Gunma 377-0702, Japan}
\altaffiltext{5} { University of Manitoba, Winnipeg, Manitoba, Canada}
\altaffiltext{6} {Osaka University, 1-1 Machikaneyama, Toyonaka, Osaka
560-0043, Japan}

\begin{abstract}
Microquasars occasionally exhibit massive jet ejections which are
distinct from the continuous or quasi-continuous weak jet ejections.
Because those massive jet ejections are rare and short events, they have
hardly been observed in X-ray so far.  In this paper, the first X-ray
observation of a massive jet ejection from the microquasar SS 433 with
the Rossi X-ray Timing Explorer (RXTE) is reported.  SS 433 undergoing a
massive ejection event shows a variety of new phenomena including a
QPO-like feature near 0.1 Hz, rapid time variability, and shot-like
activities.  The shot-like activity may be caused by the formation of a
small plasma bullet.  A massive jet may be consist of thousands of those
plasma bullets ejected from the binary system.  The size, mass, internal
energy, and kinetic energy of the bullets and the massive jet are
estimated.
\end{abstract}
\keywords{X-rays: individual (\objectname{SS 433})}

\section{Introduction}
Microquasars are stellar X-ray binaries (neutron stars or black holes)
from which relativistic jets emanate via an unknown, very efficient
mechanism \citep{mirabel99}.  Microquasars such as SS 433 and GRS
1915+105 occasionally exhibit massive jet ejections, which are
recognized as sporadic flares in their radio light curves
\citep{fiedler87,foster96}.  Because the massive jet ejections are rare
(a few per year), short (within a few days), and aperiodic, pointing
X-ray observations of these events have hardly been performed so far.
As for SS~433, no X-ray observation has been confirmed to coincide with
a radio flare, except for one or two possible coincident observations
with Einstein in 1979 \citep{band89}.  A monitoring observation over 10
days and a long-look observation lasting 13 days were performed with
ASCA in 1995 and 2000 \citep{kotani97, namiki01}, but there was no radio
flare coinciding the periods.  A multi-wavelength observation with RXTE
and the Giant Meter Radio Telescope in 2002 also misses radio flares
\citep{chakrabarti03}.  It should be stressed that the massive jets are
distinct from the stable continuous jets of SS~433 and the
quasi-continuous or weak jet of GRS~1915+105.
The radio activity of SS~433 monitored with the Green Bank
Interferometer over years may be characterized as a clustering of flare
events separated by periods of quiescent emission \citep{fiedler87}.  In
those sporadic radio flare events, the radio flux density at 2.3 GHz
exceeds 1 Jy, and massive jet blobs, which are recognized as bright
extended spots in radio images, are ejected from the core of SS 433 at a
quarter of the speed of light \citep{vermeulen93}.  The ejection of
massive jet blobs from GRS 1915+105 with a radio flux exceeding 100 mJy,
by which the source has been recognized as a microquasar in the first
place \citep{mirabel94,fender01}, have been hardly observed in X rays
\citep{muno01}, in contrast to a number of reports on the X-ray
observation of the quasi-continuous or weak jet ejections
\citep{mirabel98,klein-wolt02,ueda02}.

We report on a successful X-ray
observation of a massive jet ejection from SS~433 with the Rossi X-ray
Timing Explorer (RXTE)\@.  The observation scheme is described in
\S~\ref{sec:obs}, the data are analyzed and discussed in
\S~\ref{sec:analysis}.

\section{Observations}\label{sec:obs}
Formerly, a radio flare was the only indicator of a massive jet
ejection.  Unfortunately, an X-ray observation triggered by a radio
flare is too late to catch the moment of the ejection, as experienced in
the cases of several previous target-of-opportunity (TOO) X-ray
observations.  Because the X-ray activity precedes a radio flare, a TOO
X-ray observation will not work for a massive jet ejection event.  So we
have built a strategy to observe a {\it second}\/ massive jet ejection
event following the first event.  In the active state of SS~433, radio
flares are clustered with an interval of~$8-23$ days \citep{fiedler87}.
Therefore, a series of monitoring observations triggered by a massive
jet ejection may cover the moment of a second ejection in~23 days.

\begin{figure}
\plotone{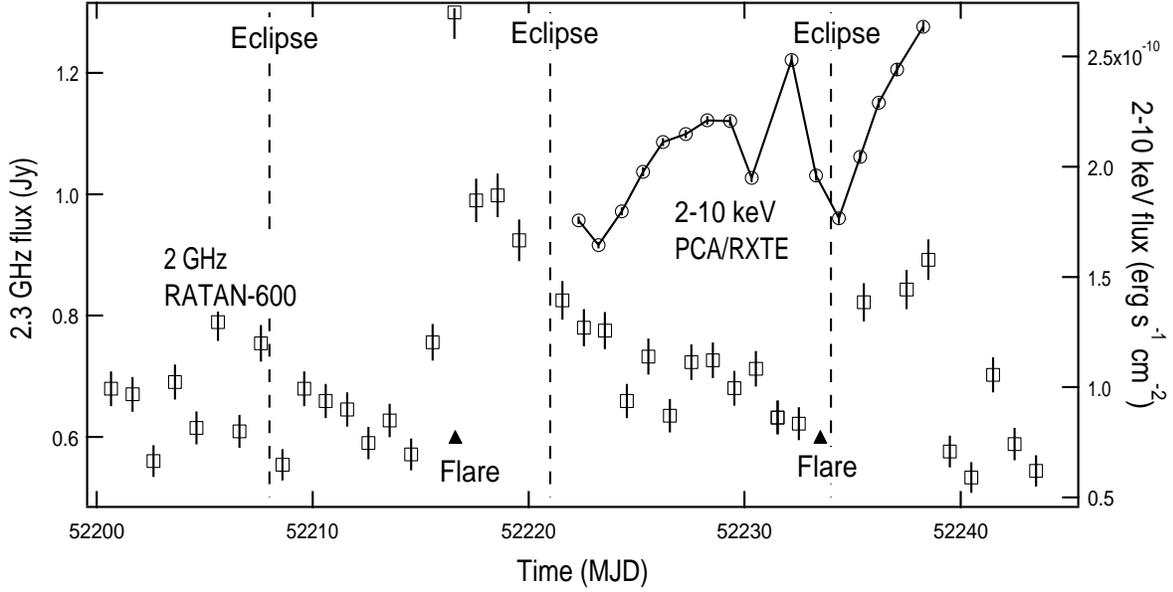} \caption{2.3 GHz radio light curve (open squares with
 error bars) and
2-10 keV X-ray light curve (open circles with
 error bars and solid line) of SS 433.
The radio flux taken with RATAN-600 shows two flares, or massive
jet-blob ejections, indicated by filled triangles.  The epochs of
eclipse are indicated in dashed lines.  The first flare has triggered
RXTE monitoring observations.  The rise of the second flare is not
prominent because of a lack of the monitoring data on MJD = 52234 due to
bad weather condition.  The X-ray fluxes show a peak just before the
second radio flare and then a dip coinciding with an eclipse.
}\label{fig:lc_x_radio}
\end{figure}

We planned a 30-days-long TOO monitoring observation of SS~433 with RXTE
to be triggered by a radio flare.  The proposal was accepted in the
Cycle 6 of the RXTE Guest Observer Program carried out for one year
beginning in March 2001.  The daily radio activity of the source has
been monitored with the RATAN-600 radio telescope \citep{korolkov79} of
the Special Astrophysical observatory of the Russian Academy of Sciences
(SAO RAS) since September 2001.  After two months of static activity
with an average flux density of 0.7 Jy at 2.3 GHz, a remarkable flare
occurred on 2001 November 2 (MJD = 52215), indicating that the source
entered its active state (Fig.~\ref{fig:lc_x_radio}).  Flux densities
reached 1.3 Jy at 2.3 GHz on MJD = 52216.6 \citep{kotani01, trushkin03}.
We started a series of X-ray observations with RXTE on MJD = 52222
\citep{kotani03, safi-harb03}.  Except for a break at MJD = 52231, SS 433
was observed for 3 ks every day.  In the X-ray light curve, a temporal
variation with time scales of $10-100$ s appeared on MJD = 52225 and the
amplitude increased day by day (Fig.~\ref{fig:lc_x}).  On MJD = 52232,
the amplitude reached a maximum, and the 2-10 keV X-ray flux reached a
local maximum of $2.5\times10^{-10}$ erg s$^{-1}$ cm$^{-2}$.  The X-ray
emission, thought to originate in the hot part of the jets as long as or
longer than $10^{12}$ cm, had never shown such a variability in past
observations \citep{safi-harb03}.  Following the maximum of the flux and
the variation amplitude, a second radio flare was detected on MJD =
52235.  Due to a missing radio data point at MJD = 52234, the precise
onset time and peak flux of the second flare are unfortunately not
known, but they are not likely out of a range $52233 < $MJD$ \leq 52235$
and $1.5$ Jy $<$ F $<$ 2 Jy.  Thus we conclude that the moment of a
massive jet ejection was observed in the X-ray band.   After the peak,
the X-ray flux dropped due to a binary eclipse.  The X-ray monitoring
observation lasted until MJD = 52238, for 17 days, providing 16 data
sets.  An observation log is shown in Table~\ref{tbl:log}.  Optical
spectroscopic observations were performed on MJD = 52220.6, 52221.6, and
52225.6 with the 0.7-m telescope at the observatory Kamenskoe Plato
\citep{mironov98}, and on MJD = 52229.39 and 52233.38 with the 0.65-m
telescope at the Gunma Astronomical Observatory
\citep{hasegawa04,kinugasa02}.  Based on the spectroscopic data, the
variation of the Doppler shifts of the jets during the campaign are
estimated.  The Doppler parameter of the receding jet is estimated to
increase from $0.07$ on MJD = 52222 to $0.13$ on 52238, and that of the
approaching jet decrease from $-0.02$ to $-0.07$.

\section{Data Analysis and Discussion}\label{sec:analysis}
\subsection{The QPO-like feature}
\begin{figure}
\plotone{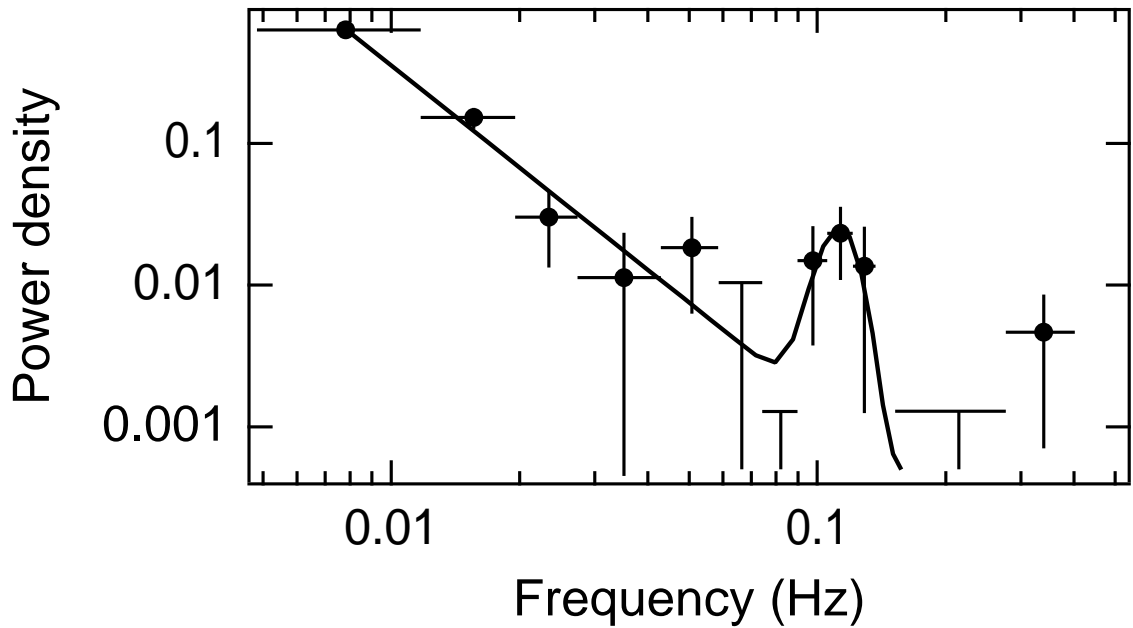}
\caption{Sum of power density spectra of the X-ray data.  The sum of all 16
spectra and the best fit model of a power-law model plus a Gaussian (solid
line) are plotted.  The data points and upper limits are marked as
filled circles and T-shaped bars, respectively.  A feature which can be
interpreted as a QPO centered at $0.1127 \pm 0.0072$ Hz with a Gaussian
sigma of $0.011 \pm 0.006$ Hz is seen.  }\label{fig:pds}
\end{figure}

Firstly, we have
searched for a periodicity in the data.  No coherent pulsation has been
detected from the 16 data sets, but a feature which can be interpreted
as a QPO has been found at 0.1 Hz in the sum of the 16 power density
spectra.  
The sum of the power density spectra
is shown in Fig.~\ref{fig:pds}.  
The fraction of the flux accounts for the QPO-like variation is
estimated from the ratio of the Gaussian normalization to the area under
the power-law continuum

This is the first detection of any
periodicity or quasi periodicity shorter than~1 day from this source.
Interestingly, other microquasars such as GRS~1915+105 also show 0.5-10
Hz low-frequency QPOs, which are considered to represent a
characteristic time scale in the accretion flow \citep{muno01}.
A super-critical accretion flow, which SS~433 is believed
to have, had not been observed in the X-ray band because of the bright
jets.  The 0.1 Hz QPO-like feature may be the first detection of the
super-critical accretion flow or disk in the X-ray band.  The
similarity to the QPO in other microquasars suggests the presence of a
common mechanism working in other systems and SS~433, at least when the
latter is undergoing a massive jet ejection.

\subsection{Spectral fitting}
The sixteen data sets have been reducted with the standard reduction
method\footnote{RXTE GOF, http://heasarc.gsfc.nasa.gov/docs/xte/xhp\_{}proc\_{}analysis.html}.  
The spectra are fitted with an empirical model,
\begin{equation}
e^{-\sigma(E) N_{\rm H}} \times \left [
		\mbox{bremsstrahlung}(kT)
		+ F_{\rm N} \times {\rm narrow line}(E_{\rm N}, \sigma_{\rm N})
		+ F_{\rm B} \times {\rm borad line} (E_{\rm B}, \sigma_{\rm B})
		\right ],
\end{equation}
where $\sigma(E)$ is the absorption cross section, $F_{\rm N}$ and
$F_{\rm B}$ are line fluxes, $E_{\rm N}$ and $E_{\rm B}$ are line center
energies, and $\sigma_{\rm N}$ and $\sigma_{\rm B}$ are line widths.
The hydrogen column density $N_{\rm H}$ and the width of the narrow line
$\sigma_{\rm N}$ are fixed to $6\times 10^{21}$ cm$^{-2}$ and 0 keV,
respectively.  The results are shown in Table~\ref{tbl:best-fit}.  The
model has been applied to SS~433's spectra obtained with the LAC/Ginga
\citep{kawai89, yuan95}, a proportional counter array whose energy
resolution and energy band are similar to those of the PCA\@.  This
model is a simple approximation of the complicated, line-abundant
spectrum revealed with finer energy resolutions of SIS/ASCA
\citep{kotani96} and HETGS/Chandra \citep{marshall02}.  In this model,
the Doppler-shifted pairs of Fe{\sc xxv} K$\alpha$, Fe{\sc xxvi}
K$\alpha$, and Ni{\sc xxvii} K$\alpha$ lines are blended into the
``narrow'' and ``broad iron lines.''  The parameters which can not be
determined from an RXTE spectrum, such as line flux ratios Fe{\sc
xxvi}/Fe {\sc xxv} and red/blue, are naturally eliminated from the
model.  The average spectrum of each data set and its evolution can be
reproduced with the model and the spectral parameters in
Table~\ref{tbl:best-fit} together with the 2-10 keV fluxes in
Fig.~\ref{fig:lc_x_radio}.

In the eclipse at MJD = 52234, both the bremsstrahlung temperature and
the line fluxes drop, as observed with the LAC/Ginga \citep{kawai89,
yuan95}.  The equivalent width of the two lines at the flux maximum (MJD
= 52232) and the eclipse (MJD = 52234) are 1.76 keV and 1.56 keV,
respectively.  The equivalent width is not sensitive to eclipse because
the base of the jet, which is responsible to both of the Doppler-shifted
line emission and the continuum emission, is occulted in eclipse
\citep{kawai89, yuan95, gies02}.

\subsection{The rapid variability}
\begin{figure}
\resizebox{!}{0.8\textheight}{\plotone{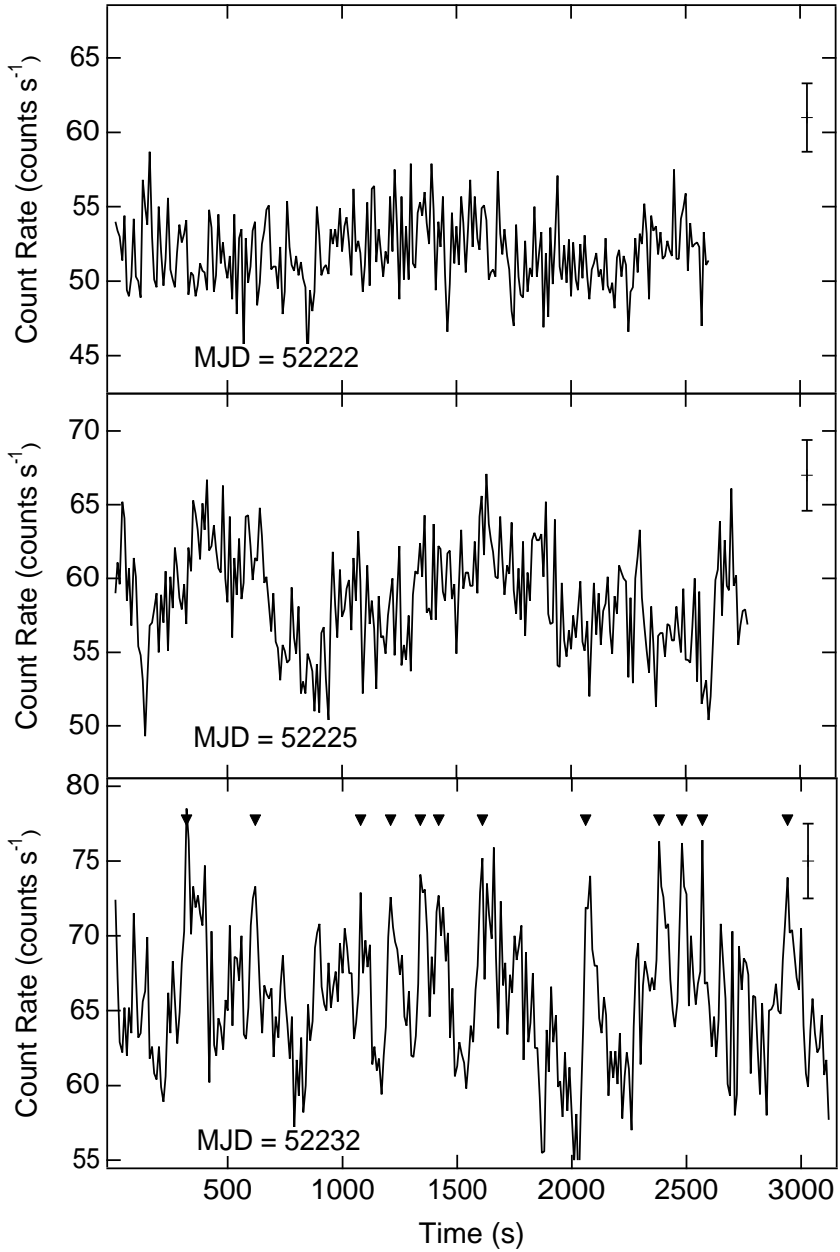}} 
\caption{Blown-up X-ray light curves taken on MJD =
52222 (upper panel), 52225 (middle), and 52232 (lower).  Only the energy
band higher than iron and nickel lines (8.38 keV for blue-shifted Ni
{\sc xxvii} K$\alpha$) is shown.  Typical 1 $\sigma$ errors are plotted
as crosses.  The amplitude of variations on MJD = 52222, if any, is not
larger than the error bar.  On MJD = 52225, the variations becomes
significant, and the amplitude reaches the maximum on 52232.  The
variation on MJD = 52232 can be interpreted as a series of ``shots'' or
``spikes.''  By picking up local maximums in the light curve above a
threshold of 74 counts s$^{-1}$ after at least 2 successive increasing
bins (30 s), twelve shots are sampled as indicated by the filled
triangles in the lower panel.  The criterion of successive increasing
bins is necessary to cut local maximums due to fluctuation.  The
threshold of 74 counts s$^{-1}$ (13 $\sigma$) is chosen so that most
shots are sampled only once.  This sampling is not exhaustive and a
different set of shots may be selected under another criterion.
}\label{fig:lc_x}
\end{figure}

\begin{figure}
\plotone{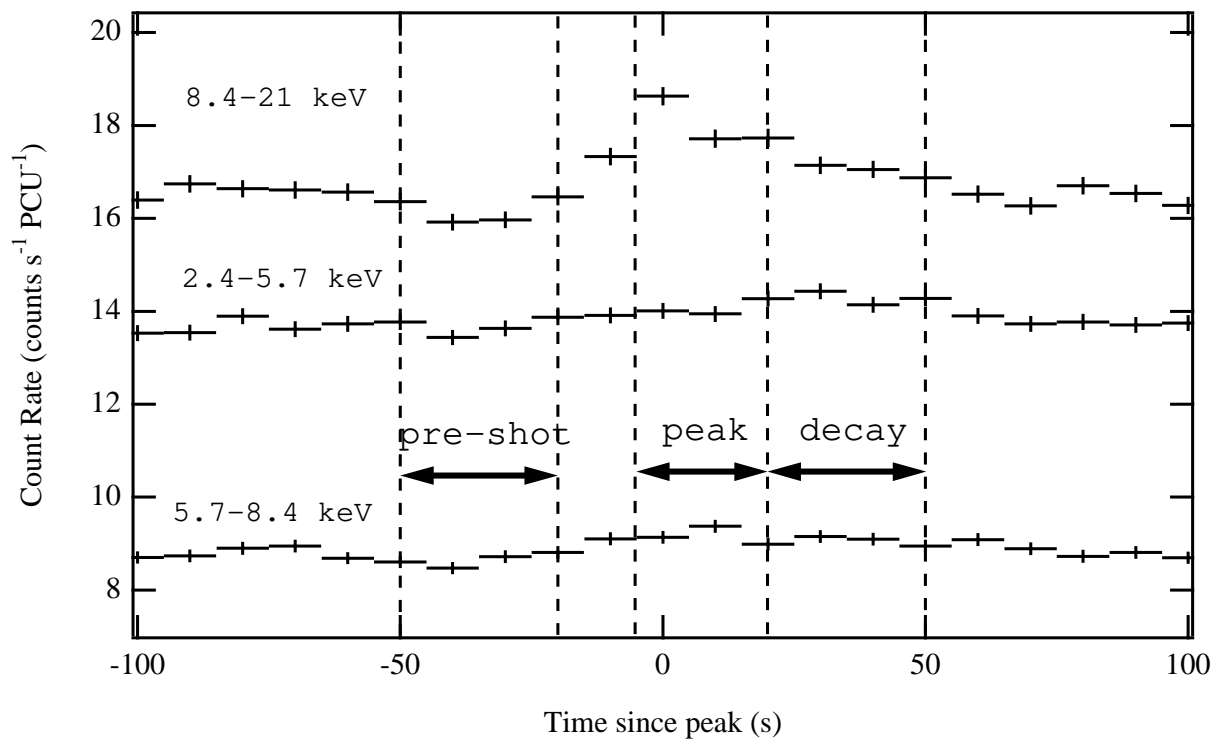}
 \caption{Average
profile of the twelve shots sampled from the data on MJD = 52232
(Fig.~\ref{fig:lc_x}).  In the 8.4-21 keV band, the
shot rises in $\sim 10$ s and decays slowly ($\sim 30$ s).  In the
 2.4-5.7 keV
band, the profile is less pronounced: The peak lags behind that in the hard band by
$\sim 30$ s, and both the rise and decay time scales are $\sim 30$ s.
The peak is indistinct in the 5.7-8.4 keV band,
too.  That  implies that the iron-line intensity is not much
 contributing the shot.
For a spectroscopic study, we have divided the profile into three phases;
the ``pre-shot,'' ``peak,'' and ``decay'' phases.  }\label{fig:foldedshot}
\end{figure}

\paragraph{Data analysis}
Then we examined the rapid variation seen on MJD = 52232
(Fig.~\ref{fig:lc_x}).  The variation, which might appear irregular or
chaotic, can be interpreted as a series of ``shots'' or ``spikes''with
widths of tens of seconds.  Their intervals are random and do not show
any periodicity.  We have sampled 12 shots as indicated in
Fig.~\ref{fig:lc_x}, and folded the light curve to make the average
profile of the shots (Fig.~\ref{fig:foldedshot}).  The shot rises fast
then slightly softens during the decay.  The 8.4-21 keV profile is
fitted with a burst model,
\begin{equation}
 \mbox{const.} +  A \times \left \{
	     \begin{array}{ll}
	      0 &   (t \leq t_0)\\
	       \frac{t-t_0}{-t_0} &(t_0 \leq t \leq 0)\\
	       \exp (-t/\tau_{\rm dec} )  & (0 \leq t)
	      \end{array}
	     \right .,
\end{equation}
where $A$ is a normalization factor, $t$ is the time from the peak,
$t_0$ is the time of the onset of the shot, and $\tau_{\rm dec}$ is the
decay time scale.  The onset time and decay time scale are fitted to be
$-23^{+5}_{-4}$ and $41_{-9}^{+12}$ s, respectively.  

\begin{figure}
\plotone{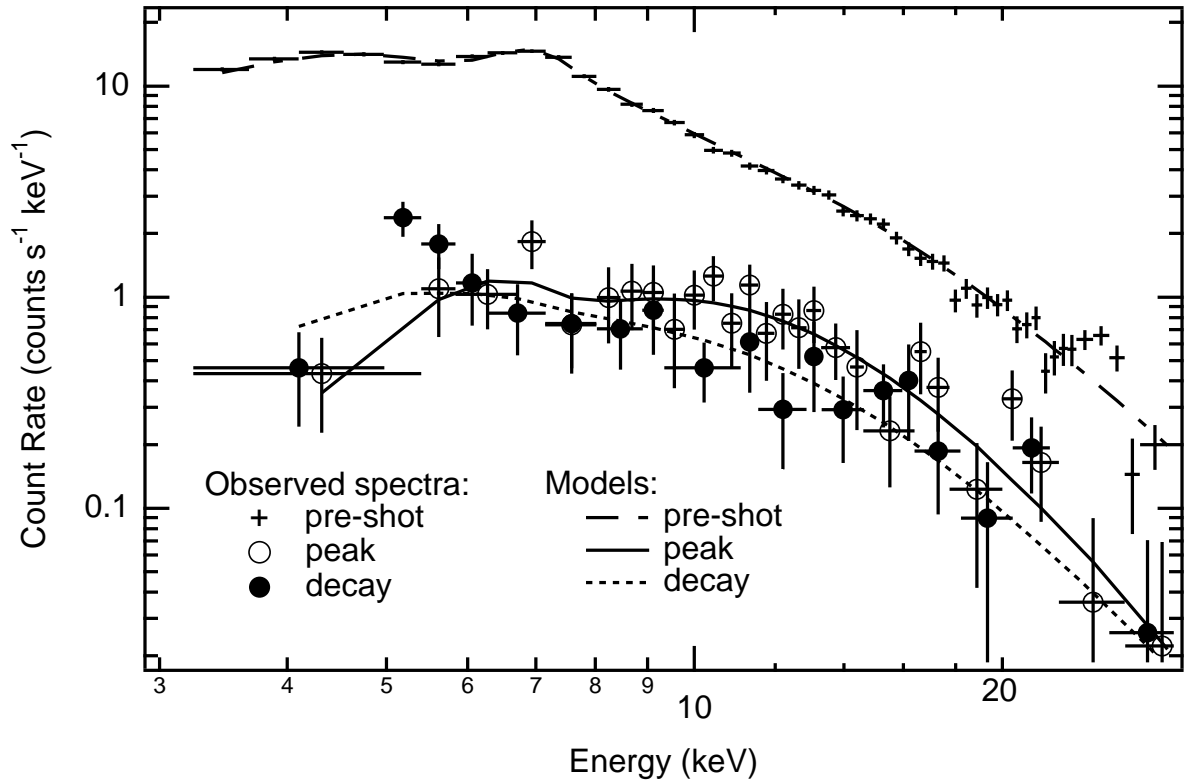}
 \caption{Spectra and best-fit models of the pre-shot phase
 (crosses), the peak phase (open circles), and the decay phase (filled
 circles).  To emphasize the spectral evolution of the shot component, 
the pre-shot spectrum is
 subtracted from the peak and decay spectra.
The background spectrum made with a tool pcabackest is
 subtracted from the pre-shot spectrum.}\label{fig:spec}
\end{figure}

We have divided the profile into three phases, namely, the ``pre-shot''
phase, the ``peak'' phase, and the ``decay'' phase, and made a spectrum
from each phase.  We have subtracted the pre-shot spectrum from each of
the peak and the decay spectra to extract the pure shot component.  The
pure shot component is shown in Fig.~\ref{fig:spec}, together with the
pre-shot spectrum.  The 3-20 keV fluxes in the peak and decay phases are
$1.9\times10^{-10}$ erg s$^{-1}$ cm$^{-2}$ and $5.7\times10^{-11}$ erg
s$^{-1}$ cm$^{-2}$, respectively.  The shot component is well fit by
either an absorbed power-law model or an absorbed thermal bremsstrahlung
model.  No emission line is detected. The total spectrum integrated over
all data taken on MJD = 52235 is expressed as an attenuated
bremsstrahlung model and require the addition of a broad iron line of
$7.00 \pm 0.02$ keV\@.  The hydrogen column density decreases in the
decay phase from $60^{+50}_{-32}\times 10^{22}$ cm$^{-2}$ to
$18^{+20}_{-13}\times 10^{22}$ cm$^{-2}$ in both models used to fit the
data.  The best fit thermal bremsstrahlung temperature and power-law
photon index at the peak are $kT_1 = 14^{+86}_{-9}$ keV and
$1.6^{+0.3}_{-0.3}$, respectively.  The index or the temperature do not
change significantly during the decay.  The unabsorbed 2-10 keV
luminosity at the peak are fitted to be $L_1 = 4.7^{+1.5}_{-1.9} \times
10^{35}$ erg s$^{-1}$ assuming a distance of $D$ = 4.85 kpc
\citep{vermeulen93}.

\paragraph{Interpretation}
This is the first detection of a rapid X-ray variability with a time
scale less than 300 s from SS~433 \citep{kotani02, kotani03,
safi-harb03}.  Although this source had been observed for numerous times
with various X-ray observatories, only variability with time scales as
long as or longer than a day had been reported.  For example, Einstein
observed the source to vary by a factor of 2 on time scales of a day
\citep{band89}, and daily variations other than the orbital and
precessional modulations are seen in Ginga and ASCA data \citep{yuan95,
kotani97}.  Temporal analysis of ROSAT data shows flickering around 3-10
s, but this variability does not appear consistently
\citep{safi-harb97}.  Since the detection with RXTE in 2001
\citep{kotani02, kotani03, safi-harb03}, evidences of rapid X-ray
variability have been accumulated.  \citet{chakrabarti03} report on
X-ray variability with time scales of a few minutes detected in 2002
with RXTE, and \citet{revnivtsev04} detected a significant X-ray
variability with time scales as short as 100 s with RXTE in 2004.

The absence or weakness of a rapid X-ray variability had been explained
in terms of the X-ray-emitting jet as long as or longer than $10^{12}$
cm.  Together with the QPO-like feature in the power density spectrum,
this shot-like variability implies the presence of X-ray sources smaller
than $10^{12}$ cm in the system.  Considering that these shots coincide a
massive jet ejection event, we further assume that they are related to
the ejection in the following discussion.

Since the spectral fit is consistent with a decrease of the absorption
hydrogen column density during the evolution of the shots, we attribute
the rise of the shots to the decrease of attenuating matter, or in other
words, the emergence of an X-ray-emitting small plasma bullet from
behind attenuating matter. Each shot literally corresponds to a shot of
a small plasma bullet from the nozzle.  This interpretation is different
from that of the X-ray variability seen in GRS~1915+105, which is
explained in terms of the rapid removal and replenishment of matter
forming the inner part of an accretion disk \citep{belloni97}.  Since
both thermal and non-thermal spectral models are consistent with the
observed spectrum, it is difficult to determine the emission mechanism.
But in either case, physical quantities of the emitting bullets would be
derived as follows.

\paragraph{Thin-thermal emission}
Given a spherical, thin-thermal, freely expanding plasma bullet with a
temperature $T(t)$, a radius $R(t) = v_{\rm exp}t$, an expanding
velocity $v_{\rm exp} =$ const., and a total number of electrons $N_{\rm
e} =$ const., the cooling would be governed with the equation
\begin{eqnarray}
 \frac 3 2 (1+X) N_{\rm e} k_{\rm B} \frac {dT} {dt} 
&= &- \frac {\Lambda(T) X N_{\rm e}^2} {\frac 4 3 \pi R^3} 
- 3 (\gamma -1) \frac 3 2 (1+X) N_{\rm e} k_{\rm B}  \frac {T v_{\rm exp}} {R}, 
\label{eq:cooling}
\end{eqnarray}
where $X = N_{\rm i}/N_{\rm e}$ is the ratio of the total number of ions
to that of electrons assumed to be 0.93117, $k_{\rm B}$ is the Boltzmann
constant, and $\gamma$ is the adiabatic index, assumed to be 5/3.  The
first term on the right-hand side corresponds to radiative cooling, and
the coefficient $\Lambda(T)$ is defined so that $\Lambda(T) N_{\rm e}/(4
\pi R^3/3) N_{\rm i}/(4 \pi R^3/3)$ equals to the emitted power per unit
volume.  The last term corresponds to expansion: For expanding plasma
with a volume $V(t)$,
\begin{equation}
 V^{\gamma-1}dT = -T(\gamma-1)V^{\gamma-2}dV \label{eq:adiabatic}
\end{equation}
is derived from the relation $TV^{\gamma-1} =$ constant.  Substituting $V
= 4 \pi R^3/3$ to Eq.~(\ref{eq:adiabatic}), the cooling rate by expansion,
$ dT/dt = -3 (\gamma-1) T v_{\rm exp}/R$
is obtained, which is equivalent to the last term of
Eq.~(\ref{eq:cooling}).
The time parameter $t$ is defined so that $t = 0$ at $R =
0$, although the radius can never be zero.  If the coefficient
$\Lambda(T)$ is proportional to $\sqrt T$ and written in the form
$\Lambda(T) = \Lambda_T \sqrt T$, Eq.~(\ref{eq:cooling}) has an analytic
solution
 \begin{eqnarray}
  \sqrt T & = & 
  \left (\sqrt {T_1} - \frac {\Lambda_T N_{\rm e}X }{4 \pi (\frac 7 2 - \frac 3 2 \gamma) (1+X) k_{\rm B} v_{\rm exp}R_1^2} \right ) 
  \left ( \frac R {R_1} \right )^{- \frac 3 2 \gamma + \frac 3 2} \nonumber \\
  &&\mbox{} + \frac {\Lambda_T X N_{\rm e}}{4 \pi (\frac 7 2 - \frac 3 2 \gamma) (1+X) k_{\rm B} v_{\rm exp}R_1^2}
  \left ( \frac R {R_1} \right )^{-2} \nonumber \\
  & = &   \sqrt {T_1} \left [\left ( 1 - \frac {L_1 \tau_{\rm exp}}{2 E_1} \right ) \left ( \frac R {R_1} \right )^{-1}
   + \frac {L_1 \tau_{\rm exp}}{2 E_1}
  \left ( \frac R {R_1} \right )^{-2} \right ],\label{eq:sol}
\end{eqnarray}
\begin{eqnarray}
E_1 &= &\frac 3 2 (1+X) N_{\rm e}  k_{\rm B} T_1\\ 
L_1 &= &\frac {3\Lambda_T \sqrt {T_1} X N_{\rm e}^2}{4\pi R_1^3}\\
\tau_{\rm exp} &=  &R_1/v_{\rm exp}
\end{eqnarray}
where the subscript ``$_1$'' denotes the value at the peak.  $E_1$, $L_1$,
 and $\tau_{\rm exp}$  correspond to the thermal energy,
the
luminosity of the bullet at the peak, and the time scale of expansion, respectively.  The ratio ${L_1 \tau_{\rm exp}}/ (2E_1)$
represents the fraction of the thermal energy in the
bullet lost by radiation (cf.\ Kotani et al.\ 1996). 
Using Eq.~(\ref{eq:sol}), the luminosity of
the bullet can be written as
\begin{eqnarray}
 L & = &  
   L_1 \left [\left ( 1 - \frac {L_1 \tau_{\rm exp}}{2 E_1} \right ) 
      \left ( \frac R {R_1} \right )^{-4}
   + \frac {L_1 \tau_{\rm exp}}{2 E_1}
  \left ( \frac R {R_1} \right )^{-5} \right ].\label{eq:lum}
\end{eqnarray}

The  expansion velocity $v_{\rm exp}$ is estimated from  the observed
temperature as
$v_{\rm exp} =  \sqrt{ k_{\rm B}T_1/(\mu m_{\rm H})}
                = 1.5^{+2.5}_{-0.6}\times10^{8}$ cm s$^{-1}$,
where $m_{\rm H}$ is the mass of
a hydrogen atom  and $\mu$ is the mean molecular weight assumed to be 0.587922.
In 40 s, the luminosity of the plasma decreases by a factor of $1/e$.  Substituting the factor into Eq.~(\ref{eq:lum}), we obtain
\begin{eqnarray}
 \frac {L_2}{L_1} = \frac 1 e & = &
   \left ( 1 - \frac {L_1 \tau_{\rm exp}}{2 E_1} \right ) 
      \left ( \frac {R_2} {R_1} \right )^{-4}
   + \frac {L_1 \tau_{\rm exp}}{2 E_1}
  \left ( \frac {R_2} {R_1} \right )^{-5}, \label{eq:l2l1}
\end{eqnarray}
where the subscript ``$_2$'' denotes the value at the decay phase.
Because the ratio of the radiation loss to the internal energy ${L_1
\tau_{\rm exp}}/ (2E_1)$ is between 0 and 1, the expansion in 40 s is
constrained as $\exp[1/5] < R_2/R_1 < \exp[1/4]$ from the above
equation.  Substituting $R_2 = R_1 + v_{\rm exp} \times 40$ s, we obtain
the radius as $R_1 = 2.2^{+3.6}_{-0.9} \times 10^{10}$ cm.  From this
radius and an observed quantity $L_1$, all other parameters are derived;
$1.06 \times 10^2 < \tau_{\rm exp} < 1.35 \times 10^2$ s, the number of
electrons $N_{\rm e} = 1.0^{+2.1}_{-0.5} \times 10^{45}$, the number
density of electrons $n_{\rm e1} = 4.1^{+9.2}_{-3.7} \times 10^{13}$
cm$^{-3}$, and the thermal energy $E_1 = 0.7^{+13.7}_{-0.6}\times
10^{38}$ erg.  Assuming that the bullet is moving at 0.26 c, the kinetic
energy of the bullet is estimated to be $0.6^{+1.2}_{-0.3}\times
10^{41}$ erg.  The assumption is consistent with the observed rise time
of 20 s, which is naturally explained by the time in which a bullet
appears from a nozzle.
Eq.~(\ref{eq:sol}) is written with these estimates as 
 \begin{eqnarray}
  \sqrt T 
  & = &   \sqrt {14 \; \mbox{[keV]}} \left [0.3 \left ( \frac R {1.6\times 10^{10} \; \mbox{[cm]}} \right )^{-1}
   + 0.7  \left ( \frac R {1.6\times 10^{10} \; \mbox{[cm]}} \right )^{-2} \right ],\label{eq:solnum}
\end{eqnarray}
where errors are omitted.

Since a shot and the unmodulated component coexist, the bullets and the
continuously emanating jet may coexist.  In that case, the small plasma
bullets can be interpreted as bright knots in the continuous jet.  The
knots are created when the mass outflow rate or the density of the
continuous jet is temporarily increased.  A temporal increase of
temperature is not plausible, because it would result in a change of the
spectrum, which has not been observed.  The properties of the knots
would not be much different from those of the small plasma bullets
discussed above, and the estimates above are valid if the knots or
bullets coexist with the continuous jet.

\paragraph{Synchrotron emission}
Because the spectral shape does not much change in the decay and because
no iron line is detected in the shot component, a non-thermal emission
from expanding bullets also can account for the shot component.  As for
the steady non-variable component, it is definitely a thin-thermal
emission with Doppler-shifted iron lines.  Therefore, it is natural to
interpret the shot component as a thermal emission, and an
interpretation of non-thermal emission is rather eccentric.  In the
following paragraph, we show physical parameters of a plasma bullet assuming that
the bullets emit X ray via synchrotron radiation.  

A power-law distribution of synchrotron electrons,
\begin{eqnarray}
 f(\gamma)d\gamma  & \equiv &
	\left\{
	 \begin{array}{ll}
	   \frac{-p+1}{\gamma_{\rm max}^{-p+1}-1} n_{\rm e,syn}  \gamma^{-p} d\gamma
	   \approx (p-1) n_{\rm e} \gamma^{-p} d\gamma
	   & (1 < \gamma \le \gamma_{\rm max})\\
	  0 & (\gamma_{\rm max} < \gamma)
	 \end{array}
	\right.
\end{eqnarray}
is assumed, where $\gamma$ is Lorentz factor of electrons, $n_{\rm
e,syn}$ is the synchrotron-electron number density, and $p =
2.2^{+1.4}_{-1.0}$ is the electron energy index derived from the photon
index $\Gamma = (p+1)/2$.  The maximum Lorentz factor should be at least
$\gamma_{\rm max} > 1.4 \times10^{5}$ to account for the X-ray emission
up to 10 keV\@.  Optically thin synchrotron flux from such a sphere is
expressed as
\begin{eqnarray}
 F_\nu &= & \frac{\chi(p)}{4\pi}
      n_{\rm e,syn} \frac{e^3}{mc^2} B^{p/2+1/2}
   \left ( \frac{4\pi mc\nu} {3e} \right )^{-p/2+1/2} \frac{4\pi R^3/3}{D^2}\label{eq:intenapprothin}\\
\chi(p) & \equiv &\frac{3^{1/2}2^{p/2-1/2}(p-1)}{(p+1)}
 \frac {\Gamma \left(\frac p 4 + \frac {19}{12}\right)
    \Gamma \left(\frac p 4 - \frac {1}{12}\right)
    \Gamma \left(\frac p 4 + \frac 5 4 \right)}
    {\Gamma \left(\frac p 4 + \frac 7 4 \right)}
\end{eqnarray}
where $\nu$ is frequency, and $B$ is the magnetic field strength in the
plasma, (e.g., \cite{hjellming95}).  By substituting the observed flux
$F\nu$(1 keV) = $1.0^{+1.0}_{-0.5}\times 10^{-2}$ photons cm$^{-2}$
s$^{-1}$ keV$^{-1}$ into Eq.~(\ref{eq:intenapprothin}), the magnetic
field strength $B$ and the total number of electrons in a bullet $N_{\rm
e}$are constrained as
\begin{eqnarray}
 \log_{10}\left[N_{\rm e}B^{p/2+1/2}\;{\rm [G]}\right]
& = & 44.6^{+3.0}_{-2.9},
\end{eqnarray}
where the rather large uncertainty is due to the uncertainty of the
electron energy index $p$.

As
the bullet expands, each high-energy electron loses energy as $E =
(R/R_1)^{-1}$ and the magnetic field and the luminosity decreases as $B
= B_1 \times (R/R_1)^{-2}$ and $L = L_1 (R/R_1)^{-2p}$, respectively.
Radiative cooling and heating are neglected.  Thus a decrease of flux by
a factor of $1/e$ corresponds to an adiabatic expansion by
$1.26^{+0.25}_{-0.10}$ of the radiating bullet.  Further assuming that
the bullet is proceeding at 0.26 $c$ with an expanding half angle of
$2.1^\circ$, which are the same value as the velocity and the half
opening angle of the continuous jet \citep{namiki03}, the expansion
velocity is estimated as $v_{\rm exp} = 2.9 \times 10^{8}$ cm s$^{-1}$.
From this expansion velocity and the expansion factor obtained above,
the radius of the plasma is determined as $R = 4.5^{+1.4}_{-1.4} \times
10^{10}$ cm, which is consistent with the rise time of 20 s.  

The synchrotron-electron number density and the strength of the magnetic
field can be estimated if their energies are assumed to be in
equipartition, i.e.,
\begin{eqnarray}
 \frac {B^2}{8\pi} & = & \int_1^{\gamma_{\rm max}} \frac{p-1}{\gamma_{\rm max}^{-p+1}-1}
n_{\rm e} mc^2 \gamma^{-p+1} d\gamma \\
 &&\approx \left\{ 
	    \begin{array}{ll}
	     \frac{p-1}{-p+2}\gamma_{\rm max}^{-p+2} n_{\rm e}mc^2 & \;\;\;\; (1 < p < 2)\\
	     \frac{p-1}{p-2} n_{\rm e}mc^2 & \;\;\;\; (2 < p).
	    \end{array}
	   \right. \label{eq:enee}
\end{eqnarray}
Substituting an electron energy index of $p=2.2$, we obtain $B = 1.8
\times 10^{2}$ G and $n_{\rm e} = 2.4 \times 10^{8}$ cm$^{-3}$. The
total number of electrons and internal energy in a bullet are estimated
as $N_{\rm e} = 9 \times 10^{40}$ per shot and $E_{\rm syn} = 5.3 \times
10^{35}$ erg, respectively.  These number of electrons and internal
energy derived here are smaller by orders of magnitude than those of the
thin-thermal case.

\paragraph{Comptonized emission}
If the emission mechanism is inverse-Compton scattering of optical
photons, the seed-photon density would decrease and the emission would
decay as the plasma bullet gets away from the central engine.  Based on
this Comptonized-emission model, the parameters of the plasma bullet,
such as the electron number density, the total number of electrons, and
the total energy of electrons, are estimated.  They are found to be not
much different from those in the case of synchrotron emission, although
the uncertainties of parameters are larger in the case of
inverse-Comptonization.

\subsection{The massive jet}
Based on the bullets model, we suggest an explanation of the massive jet
ejection: During the massive jet ejection event, small discrete plasma
bullets, or knots in the continuously emanating flow, are successively
ejected at random intervals of $\sim 150$ s.  The radius of bullets is
estimated as $R_1 = 2.2^{+3.6}_{-0.9} \times 10^{10}$ cm.  The X-ray
emission from the small plasma bullets, either thermal or non-thermal,
decays in 40 s as it expands.  Assuming that the most active state lasts
3 days, the total number of small bullets ejected in a single massive
jet event is estimated to be 1700.  The total mass and total kinetic
energy of all the 1700 bullets are $3.3^{+6.7}_{-1.8} \times 10^{24}$ g
and $1.0^{+2.1}_{-0.5} \times 10^{44}$ erg, respectively.  The average
mass ejection rate and average kinetic luminosity over 3 days are
$0.7_{-0.4}^{+1.6} \times 10^{16}$ g s$^{-1}$ and $3.9^{+8.0}_{-2.1}
\times 10^{38}$ erg s$^{-1}$, respectively.  In the case of synchrotron
emission from baryonic plasma, only the lower limits of mass and kinetic
energy are derived; the total mass and total kinetic energy of the 1700
bullets would be at least $3.0 \times 10^{20}$ g and $9.1 \times
10^{39}$ erg, respectively, and the average mass ejection rate and
average kinetic luminosity would be at least $1.1 \times 10^{15}$ g and
$3.5 \times 10^{34}$ erg s$^{-1}$, respectively.

The estimated average kinetic luminosity of $\sim 10^{38}$ erg s$^{-1}$
is considerably lower than estimates based on the quiescent or normal
state.  For example, \citet{kotani97} has calculated the kinetic
luminosity as $1 \times 10^{40}$ erg s$^{-1}$ based on ASCA data,
\citet{marshall02} as $3.2 \times 10^{38}$ erg s$^{-1}$ based on
HETGS/Chandra data, and \citet{brinkmann05} as $5 \times 10^{39}$ erg
s$^{-1}$ based on EPIC/XMM-Newton data.  It is puzzling that the mass
outflow rate and kinetic luminosity in the massive jet ejection are not
so ``massive'' compared to those of the steady continuous jet flow seen
in most occasions.  There are several possibilities to account for the
inconsistency in terms of the bullets model: 1) The mass outflow rate
and kinetic luminosity of a massive jet are not larger than those of
quiescent steady jet, but the efficiency to accelerate electrons
contributing to synchrotron radio emission is far larger. 2) The massive
jet is not an assembly of the small plasma bullets, but mainly supplied
with the steady flow which coexists with the bullets.  3) In spite of
the monitoring observation with a sampling rate of 3 ks a day, we have
missed the moment of the true massive jet ejection, which lasts only,
say, hours, and a massive jet of $10^{44} - 10^{45}$ erg is ejected at a
maximum outflow rate of $10^{40} - 10^{41}$ erg s$^{-1}$.  We do not yet
have an evidence supporting one of them, but suggest that the second
case is unlikely, in which the unvariable X-ray component is expected to
rise as the radio flux densities rise.

Another question is whether X-ray variability is really related to radio
flaring, which is associated with blob rebrightening events out of the
system core.  At 1.6 GHz, radio flares peak at 35 AU from the core
\citep{paragi99}.  If radio flares were caused by an environmental
condition, it would not be detectable in X-ray band at the ejection, and
the coincidence of the rapid X-ray variability and the massive jet event
would be accidental.  However, \citet{paragi99} suggest that the
rebrightening is due to the attenuation by out-flowing gas around the
core.  If the cause of radio flares is not an environmental condition
but a core activity, the activity which changes the radio flux by a
factor of 2 might be detectable in X-ray band.  That should be tested in
future multi-wavelength observations.

X-ray shots are firstly seen in the data from MJD = 52225, a week before
its maximum activity and the onset of a second radio flare.  Provided
that shots precede a radio flare, we can predict a massive
jet-ejection event based on X-ray monitoring data.  On detection of
X-ray shots, notice of a massive jet ejection to occur in a week can be
sent to ground and space observatories.  An observation campaign
covering a massive jet ejection is possible.  There is still a
possibility that the duration of a massive jet ejection is shorter
than~1 day and the moment has been missed even in our observations.  It
can be confirmed in future observations.  And this technique may be
applicable for prediction of massive jet ejections from other
microquasars.  In spite of numerous observations performed so far, it is
not yet known whether massive jet ejections from other microquasars such
as GRS~1915+105 are also preceded by a precursor or not.  As a specially
coordinated observation is required to detect the shot-like variability
from SS~433, a carefully coordinated observations plan is desirable to
observe GRS~1915+105 in a massive jet ejection event with an X-ray
mission.  The findings reported here imply that new and important
physics of a microquasar is revealed by observing massive jet ejections.
The observation of these events is essential to explore the nature of
microquasars.  Therefore the technique to observe massive jet ejections
is one of the most important results from this study.  Future
observations of massive jet ejections from microquasars are encouraged.


\acknowledgments
T.~K. is supported by a 21st Century COE Program at Tokyo Tech
``Nanometer-Scale Quantum Physics'' by the Ministry of Education,
Culture, Sports, Science and Technology.  S.~A.~T. is very grateful to
the Russian Base Researches Foundation for support of the SS~433
monitoring with the RATAN-600 radio telescope.  K.~K. is supported by a
grant-in aid 16740121 from the Ministry of Education, Culture, Sports,
Science and Technology of Japan.  We thank the anonymous referee for
useful comments and suggestions to improve this paper.

\clearpage

\begin{table}
\centering
\caption{Observation log}\label{tbl:log}
\begin{tabular}{cccl}
\hline
Start &End &Exposure Time &PCU$^{\rm a}$\\
 (MJD) &  (MJD) &(ks) &\\
\hline
2001/11/09 07:10 (52222.299) &2001/11/09 08:11 (52222.341) &3.6&0234 \\
2001/11/10 05:19 (52223.222) &2001/11/10 06:21 (52223.265) &3.7&0234 \\
2001/11/11 06:47 (52224.283) &2001/11/11 07:47 (52224.325) &3.6&0234 \\
2001/11/12 06:35 (52225.275) &2001/11/12 07:35 (52225.316) &3.5&0234 \\
2001/11/13 04:46 (52226.199) &2001/11/13 05:46 (52226.240) &3.5&0234 \\
2001/11/14 06:11 (52227.258) &2001/11/14 07:13 (52227.301) &3.7&0234 \\ 
2001/11/15 06:00 (52228.250) &2001/11/15 07:01 (52228.293) &3.7&0234 \\ 
2001/11/16 07:26 (52229.310) &2001/11/16 08:32 (52229.356) &4.0&023  \\
2001/11/17 07:14 (52230.302) &2001/11/17 08:20 (52230.349) &4.1&023  \\
2001/11/19 03:41 (52232.154) &2001/11/19 04:33 (52232.190) &3.1&0234 \\
2001/11/20 06:40 (52233.278) &2001/11/20 07:46 (52233.324) &4.0&02   \\ 
2001/11/21 08:05 (52234.337) &2001/11/21 09:16 (52234.387) &4.3&023  \\ 
2001/11/22 07:54 (52235.330) &2001/11/22 09:05 (52235.379) &4.2&024  \\
2001/11/23 04:32 (52236.189) &2001/11/23 05:29 (52236.229) &3.5&024  \\
2001/11/24 01:11 (52237.050) &2001/11/24 01:54 (52237.080) &2.6&02   \\
2001/11/25 05:45 (52238.240) &2001/11/25 06:47 (52238.283) &3.7&012  \\
\hline
\end{tabular}
$^{\rm a}$ Proportional counter units in operation.
\end{table}

\clearpage

\begin{table}
\caption{Spectral parameters}\label{tbl:best-fit}
\begin{tabular}{ccccccccc}
\hline
Start     &$kT$ &$E_{\rm N}$&$F_{\rm N}$            &$E_{\rm B}$&$\sigma_{\rm B}$&$F_{\rm B}$            &$\chi^2/$d.o.f.\\
(MJD)     &(keV)&(keV)      &(ph s$^{-1}$ cm$^{-2}$)&(keV)      &(keV)           &(ph s$^{-1}$ cm$^{-2}$)&\\
\hline
52222.299 &35   &6.5        &$5.1\times10^{-4}$     &6.92       &0.83            &$2.2\times10^{-3}$     &30.5/31\\
52223.222 &49   &6.4        &$4.7\times10^{-4}$     &6.98       &0.81            &$2.2\times10^{-3}$     &51.5/31\\
52224.283 &47   &6.5        &$4.7\times10^{-4}$     &6.98       &0.83            &$2.2\times10^{-3}$     &29.7/31\\
52225.275 &56   &6.7        &$5.2\times10^{-4}$     &6.95       &1.00            &$2.9\times10^{-3}$     &57.0/31\\
52226.199 &55   &6.8        &$6.5\times10^{-4}$     &6.90       &1.07            &$3.1\times10^{-3}$     &82.8/31\\
52227.258 &45   &6.9        &$5.5\times10^{-4}$     &6.90       &1.10            &$3.1\times10^{-3}$     &45.0/31\\
52228.250 &37   &6.8        &$4.8\times10^{-4}$     &6.92       &0.99            &$2.8\times10^{-3}$     &39.9/31\\
52229.310 &48   &6.9        &$6.3\times10^{-4}$     &6.94       &1.07            &$2.9\times10^{-3}$     &38.7/31\\
52230.302 &37   &6.9        &$3.4\times10^{-4}$     &6.97       &0.95            &$2.3\times10^{-3}$     &49.6/31\\
52232.154 &38   &7.0        &$5.9\times10^{-4}$     &6.88       &1.21            &$3.3\times10^{-3}$     &72.2/31\\
52233.278 &25   &7.0        &$2.2\times10^{-4}$     &7.06       &0.93            &$2.0\times10^{-3}$     &87.1/30\\
52234.337 &21   &6.9        &$2.2\times10^{-4}$     &7.11       &1.07            &$2.1\times10^{-3}$     &42.1/30\\
52235.330 &48   &7.0        &$6.0\times10^{-4}$     &6.88       &1.12            &$2.5\times10^{-3}$     &38.4/30\\
52236.189 &53   &7.0        &$5.1\times10^{-4}$     &6.92       &1.05            &$2.7\times10^{-3}$     &53.1/30\\
52237.050 &55   &7.0        &$6.6\times10^{-4}$     &6.74       &1.24            &$3.0\times10^{-3}$     &51.9/30\\
52238.240 &42   &7.1        &$6.9\times10^{-4}$     &6.92       &1.27            &$3.4\times10^{-3}$     &42.9/30\\
\hline
\end{tabular}
The least significant digit of each best-fit value has an uncertainty of
the same order.  For 2-10 keV flux, see  Fig.~\ref{fig:lc_x_radio}.
\end{table}

\clearpage

\clearpage

\clearpage

\end{document}